# *Anomalicious*: Automated Detection of **Anomal**ous and Potentially Mal**icious** Commits on GitHub

Danielle Gonzalez  
Rochester Institute of Technology  
dng2551@rit.edu

Thomas Zimmermann, Patrice Godefroid  
Microsoft Research  
tzimmer@microsoft.com  
pg@microsoft.com

Max Schäfer  
GitHub  
max-schaefer@github.com

*Abstract*—Security is critical to the adoption of open source software (OSS), yet few automated solutions currently exist to help detect and prevent *malicious contributions* from infecting open source repositories. On GitHub, a primary host of OSS, repositories contain not only code but also a wealth of commit-related and contextual metadata – *what if this metadata could be used to automatically identify malicious OSS contributions?*

In this work, we show how to use only commit logs and repository metadata to automatically detect anomalous and potentially malicious commits. We identify and evaluate several relevant factors which can be automatically computed from this data, such as the modification of sensitive files, outlier change properties, or a lack of trust in the commit's author. Our tool, *Anomalicious*, automatically computes these factors and considers them holistically using a rule-based decision model. In an evaluation on a data set of 15 malware-infected repositories, *Anomalicious* showed promising results and identified 53.33% of malicious commits, while flagging less than 1% of commits for most repositories. Additionally, the tool found other interesting anomalies that are not related to malicious commits in an analysis of repositories with no known malicious commits.

## I. Introduction

Modern software development heavily relies on third-party libraries. Developers share and find them using registries such as the Node Package Manager (NPM), which hosts JavaScript packages and, with over 1 million packages, is one of the largest registries in the world. Supply-chain attacks target development ecosystems and package repositories [1], [2], and in recent years many have succeeded in injecting malware into package managers [3], [4]. This poses a significant security risk for developers and end users, as dependence on malicious packages will compromise their software supply chain.

Protecting against supply-chain attacks is difficult because many attacks are possible. For example, malicious dependencies can be introduced to a project from a compromised ("hijacked") account [5] or via typo squatting, which preys on minor typographical mistakes in package names by developers [6]. Another attack pattern has an attacker commit malicious code to a well-known package that many others depend on. A recent example is the popular `event-stream` repository (1.9 million weekly downloads on NPM), which infamously fell victim to a back-door injection by a contributor granted elevated privileges by the repository owner [7].

In this paper, we focus on attacks that make malicious contributions via commits to open source repositories. To address challenges in preventing such attacks, we've developed a tool, *Anomalicious*, that automatically detects and flags **anomal**ous and potentially mal**icious** commits (Section III-D). *Anomalicious* mines commit and repository metadata to learn a project's history, using this data to compute several security-related factors for each commit. A rule-based decision model is used to compare factor values for an individual commit to threshold values based on "normal" values learned from the history and other settings users can customize to suit their individual needs. Based on this analysis, *Anomalicious* flags commits exhibiting suspicious properties and produces a detailed report to explain why they were flagged.

A successful commit-anomaly detector should satisfy five criteria: (1) *language agnostic*, to accommodate projects using any (or multiple) languages, (2) *customizable*, to suit individual needs and concerns, (3) produce *explainable* alerts, so developers can take informed action, and (4) make *infrequent* alerts (low positive rate), as anomalies are expected to be rare by definition [8]. Finally, a *security-oriented* anomaly detector such as ours must be able to (5) *detect malicious commits*. The first 3 criteria are directly addressed in the tool's design by using language-agnostic data and a rule-based decision model allowing customization and explanations. We show through experiments that *Anomalicious* succeeds in having a low positive rate and in detecting malicious commits (Section V).

The main contribution (Section VI-A) of this work is showing that **malicious commits can be detected with high accuracy using a security-focused rule-based anomaly detector relying only on commit-related metadata**.

## II. Background and Related Work

Before discussing *Anomalicious*' design and evaluation, we discuss the context and motivation for this work. Many software products are developed using an open-source model, where code and other artifacts are hosted publicly on a repository platform like GitHub so anyone interested can contribute. While there are many benefits to this, **OSS projects face an increased risk of being compromised by malicious contributions**. Of particular concern are *supply-chain attacks* [1], [2], which inject malware into the code of popular software packages in order to infect the supply chain of other systems depending on these. The Open Web Application Security Project (OWASP) lists *"Using Components with Known Vulnerabilities"* as one of the Top 10 Web Application Security Risks [9], and findings from prior studies illustrate the scope and impact of these attacks. There are several types of attack, such as typo squatting, *i.e.* naming malicious packages with slightly-misspelled names of popular packages or, most relevant to this work, injecting malicious code into existing packages [3]. The impact of these attacks are severe: **adding a single package dramatically increases a system's attack surface due to the "nested" nature of dependencies** [10], [11]. Duan *et al.* [4] built a "vetting pipeline" that uses metadata, static, and dynamic analysis to scan registries for suspicious packages; they identified 339 malicious packages, of which 82% were previously unknown. This effort aligns with ours, but we aim to *prevent* malicious injections and use only repository and commit metadata available from GitHub.

**There are few automated solutions to help maintainers detect and prevent malicious contributions**, especially those affecting parts of the *software supply chain* [1] other than source code. *Anomaly detection* is a promising technique that could be used for this purpose: something is anomalous when it is *"inconsistent with or deviating from what is usual, normal, or expected"* [8]. Anomaly detectors monitor activities within an environment (*e.g.* repository) and raise *alerts* when anomalies are detected. Unusual behavior is not always malicious, but alerting maintainers to anomalies increases their *awareness* so they can quickly decide if action is needed.

**Several prior studies have demonstrated how anomaly detection techniques can be successfully applied to repository data from GitHub.** Alali *et al.* [12] did not build an anomaly detector, but used similar techniques to define a "typical" commit based on change properties *e.g.* lines of code (LOC) mined from commit logs. Three studies built and applied anomaly detectors to address the problem of notification overload [13] by detecting several types of unusual events and identifying those developers most wanted to be alerted to. Goyal *et al.* [14]'s experiment considered 10 commit factors, and found developers were more interested in alerts when they explained *why* the commit was anomalous. This supported earlier findings from Leite *et al.* [15]; their anomaly detector considered 12 unusual commit events, presenting alerts in the form of a *dashboard*. In their evaluation, they received feedback on the importance of providing developers with "justification". We have applied these findings to *Anomalicious*' flagged commit reports, as shown in Figure 3. Treude *et al.* [16]'s experiment found that developer's were interested in 6 of the 30 "unusual events" considered for commits (12), issues (6), and pull requests (12). All of these approaches used statistical outlier detection to define and identify anomalies, and while unique factor sets were considered they all found change properties to be effective for this purpose. This factor is also considered by *Anomalicious* as described in Section III-A. While not an anomaly detector, Rosen *et al.* [17] built a prediction model for "risky" commits that used 12 commit and repository factors, including change properties. ***Anomalicious'* unique rule-based decision model and security-oriented factors can identify suspicious and potentially malicious changes to code *and* development infrastructure (*e.g.* build tools).** It is also highly customizable and explainable, so users can easily configure it to suit their individual needs.

## III. COMMIT FACTORS

To identify anomalous and potentially malicious commits, *Anomalicious* mines commit logs and repository metadata (Section IV-A) to compute a set of relevant **factors**. A factor represents properties of a commit or contributor for which certain values may indicate suspicious or unusual behavior. We do not claim any *single* factor can be used to determine if a commit is anomalous or malicious. *Anomalicious* computes values for each factor individually, but a commit is only flagged if multiple factors have values violating a set of rules (Section IV-C). The tool currently considers 5 factors, described herein: (1) outlier change properties, (2) sensitive files, (3) file history, (4) pull requests, and (5) contributor trust. In Section IV, we describe how *Anomalicious* collects the relevant data, computes each factor, and evaluates their values to reach a decision.

### A. Outlier Change Properties

Identifying anomalous commits first requires understanding what a "typical" commit looks like [12]. The simplest approach is to quantify properties of a commit, measure their values over the repository's history, and use descriptive statistics to define average values for each. Commit objects, as time-stamped "snapshots" of a repository, store several types of information that could be considered. For example, an anomaly detector could measure the average time-of-day that commits are pushed to a repository; commits made at "odd" times would be flagged, where "odd" is based on a pre-defined threshold for difference from the average value.

For this factor, *Anomalicious* analyzes **seven change properties for each commit**, looking for outlier values. Change properties are quantified characteristics of the changes made to the repository's files (*e.g.* number of files added, deleted) and their contents (*e.g.* lines of code added, removed) between the commit and its predecessor/parent. Our selection criterion was simple: each change property should only require data that is easily extracted from the commit log, and have been proven effective for anomaly detection by at least one prior study. The first five factors were used in all prior studies of typical [12] or anomalous [14]–[17] commit studies: **lines of code (LOC)** (1) **added** and (2) **removed**, and **the number of files** (3) **added**, (4) **removed**, and (5) **modified**. We also consider (6) the **number of files renamed**, first used by Treude *et al.* [16], and (7) **the number of unique file *types* modified**, first used by Goyal *et al.* [14]. Considering these properties can be generated for any pair of commits, our approach computes those for each commit and its immediate parent.

*Anomalicious* uses the repository history built by the data pipeline (Section 2) to parse commit data and consider the outlier change property factor at two scopes. First, the mean value of each change property across all commits in the repository is computed to get the repository's mean change properties. Second, each contributor's mean change properties are computed from all their authored commits. Based on prior work [12], [16] and preliminary experimentation, we selected an anomaly value of two standard deviations from the mean for this factor. This anomaly value and the minimum number of outliers that must be found are customizable.

### B. Sensitive Files

Storing files containing sensitive data in a publicly accessible location is a well-known bad practice [18]. Files storing private or confidential data (*e.g.* credentials) are clearly sensitive, but it is also important to consider those which are part of the software *supply-chain* [1] like build (*e.g.* `pom.xml`) and configuration (*e.g.* `package.json`) files. These are also sensitive because they can be leveraged in a *supply-chain attack* (see Section II), for example, to inject a backdoor into the system. Their access should therefore be closely monitored to detect malicious modifications [2].

To support such monitoring, we include changes to sensitive infrastructure files as a factor in *Anomalicious*. We were first motivated to do so by the Octopus Scanner Malware [19], which affected NetBeans projects by injecting malicious payloads and modifying build files to spread itself. However, to ensure that this factor can generalize to detect other past and future attacks, we consider more than just specific build files.

In our implementation of *Anomalicious*, **any files of these types are considered potentially sensitive**: `.xml`, `.json`, `.jar`, `.ini`, `.dat`, `.cnf`, `.yml`, `.toml`, `.gradle`, `.bin`, `.config`, `.exe`, `.properties`, `.cmd`, `.build`. But this list can also be customized to consider other file types or specific file *names* instead.

### C. File History

Another useful consideration for anomaly detection is the *history* of each file in the commit, specifically the proportion of ownership each contributor has to the file, and whether the commit author has ever modified the file before. As defined by Bird *et al.* [20], a contributor's **proportion of ownership** for a file is the ratio of the file's commits authored by the contributor to the total number of commits made to the file. Furthermore, the **file owner(s)** authored the *highest* proportion of commits, and the **majority contributor(s)**' proportion of commits is above a specified threshold [20].

In the context of anomaly detection, knowledge of ownership can be used to identify commits where the author has modified files that they do not own or have not made significant contributions to in the past. We acknowledge that it may be normal for this to occur in any specific repository. However, this behavior could become suspicious when combined with other factors. For example, consider a commit whose author has made an unusually large addition (LOC) to a configuration file. This could be benign *or* suspicious, depending on *who* made this change - *e.g. why is a contributor, who has only made 1% of prior changes, adding so much to the file?*

We also note when it is a **commit author's first time modifying a file**, also used by Rosen *et al.* [17] and Leite *et al.* [15]. Similar to commit proportions, first file changes must be considered in the context of other factors because it is critical that *new contributors are not automatically deemed suspicious*. However, a commit whose author has made several past contributions but now modifies a sensitive file for the first time might be worthy of investigation.

*Anomalicious* is implemented to allow multiple contributors to have the owner role, accounting for situations such as two contributors who both authored 50% of a file's commits. Multiple majority contributors are also allowed; in our implementation the commit proportion threshold (minimum percentage of total commits) for 'majority' is configurable, but defaults to *50% or more of the file owner's commit proportion*. In other words, if the file's owner made 20% of its commits, anyone with a commit proportion between 10% and 19% would be a majority contributor. These proportions were chosen to avoid a restriction for minimum number of commits a file must have, which would occur if we used a static *number* of commits to define the roles. Further, our default threshold for majority contributor prevents a contributor from holding both roles, which makes analysis more efficient. The decision model uses rules to determine if first or atypical file changes are suspicious, as discussed in Section IV-C.

### D. Contributor Trust

Open-source projects often use a development model where new contributions must be peer-reviewed before they are merged by a "core" contributor with push access. When studying these projects, it is important to consider that these decisions are influenced by technical *and* social factors [13], [21]. One very influential social factor identified in prior work is *trust among contributors* [13], [21]–[25]. A reviewer's trust in the author can influence the rigor of evaluation [21], [25] and the decision to merge or reject a contribution [22], [24]. To address this, we've designed a technique to **label contributors as trusted or untrusted based on their contribution history for the repository being considered**. We feel this will help contextualize anomalous behavior, *e.g.* a maintainer might care less about a commit with outlier change properties and a first-time file change if the author is trusted.

There is no standard technique for quantifying trust, so we chose six **trust factors** inspired by prior work and malicious attacks: **identifiable username**, **account age**, **number of commits**, **time since first commit**, **commit-time distribution**, and **proportion of pull requests rejected**. These factors can be used to reason about a contributor's past engagement and participation in the project, and detect suspicious activity outside the scope of an individual commit. For example, a maintainer may be more interested in the activities of a contributor with a day-old account who rapidly authored several commits in that time. In Section IV-B, we describe how these values are holistically considered to infer trustworthiness.

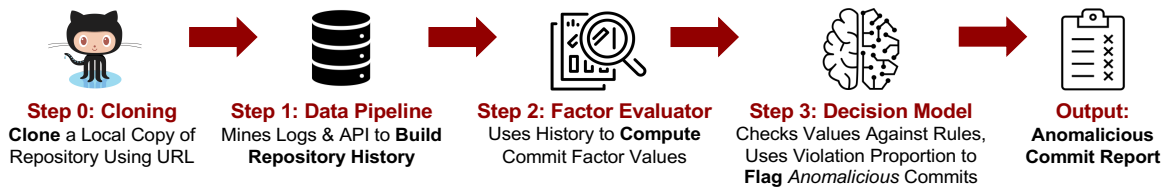

Fig. 1. Overview of Anomalicious Commit Detector Components, Data Flow, and Output

## E. Pull Requests

It is common for open source projects to use a fork-and-pull model, where only a core set of contributors have direct push permissions to the main repository, and contributions undergo peer review via pull requests (PRs). These artifacts provides a lot of contextual data for a commit such as discussions, labels, assignee, *etc*. that can be used to better understand *what* a commit is doing and *why*. Based on the other factors selected for use with *Anomalicious*, we chose to include **rejected pull requests** as a contextual factor. Specifically, *Anomalicious* analyzes all pull requests associated with a given commit to identify any whose status is "closed" and has no "merged at" date. If a commit is found to exhibit suspicious behavior, association with a rejected pull request could amplify this suspicion or provide additional context for the maintainer reviewing an anomaly alert. We also consider if contributors have rejected pull requests for our *contributor trust* factor (Section III-D). The tool can be extended to consider other scenarios such as association with a merged PR or no PR at all. These were not used for our experiments because they were not effective when analyzing collections of repositories due to varying numbers of contributors or development models used. However, the tool is implemented to retrieve all data available from the GitHub API for each pull request so additional attributes could easily be added.

## IV. TOOL DESIGN

Figure 1 is an overview of *Anomalicious*' design. There are 3 components: the **Data Pipeline**, the **Factor Evaluator**, and the **Decision Model**. When the tool is run, it produces an **Anomalicious Commit Report**. In this section, we describe each component and give an example of how the tool's findings are explained for each flagged commit in a report.

### A. Data Pipeline

The first component of *Anomalicious* is the **Data Pipeline**. This is the process in which the repository and commit data needed to compute each factor is collected, organized, and stored. An important design requirement for this component was that the relationships between data must be preserved; simply collecting and logging all filenames, commit hashes, etc. in the repository would hinder factor computation. For example, a commit will reference one or more *files*, each of which has one or more *contributors*. These connections are necessary to measure factors such as *file ownership*. *Anomalicious* stores the collected data as a series of objects which represent the **Repository History** as shown in Figure 2.

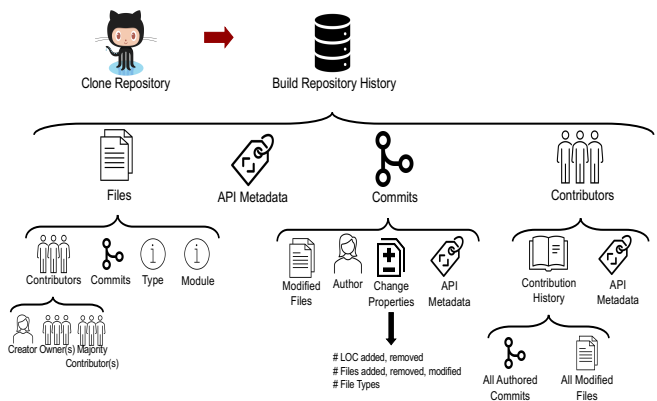

Fig. 2. Organization of Repository History Built by the Data Pipeline

Data collection begins by cloning a local copy of the repository and initializing a **Repository** object that acts as a "root" for the rest of the data. This object stores references to all files, commits, and commit authors, metadata from the GitHub API, and other repository-level information needed for the factors (*e.g.* average change properties across all commits). The `pydriller` [26] Python library was used to identify and iterate through each of the repository's commits, and *Anomalicious'* Repository and Commit objects maintain references to the corresponding objects generated by this library. We excluded merge commits from the analysis, so each commit (except the first) has a unique parent. A **Commit** object is initialized for each, which stores the descriptive details (committer, author, hash, *etc*.), change properties, and lists of file names and file types modified in the commit. For each file in the repository, a new **File** object is initialized to store references to all prior contributors, specifying the creator, owner(s), and majority contributor(s), commits, the file type, and module. Finally, the set of **Contributors** to the repository is identified by extracting the name and email of each unique commit author during commit processing. Contributor objects store data related to a specific repository as a **Contribution History** object. Each contribution history object stores a dictionary of all files modified by the contributor (with a list of commit hashes for each file) and a list of Commit objects authored by the contributor.

### B. Factor Evaluator

Once built, the Repository History is passed to the Factor Evaluator to compute values for the five commit factors defined in Section III-D. For the *outlier commit properties* factor, all modifications in the commit are parsed to aggregate the "count" (*e.g.* LOC added) for each change property. These aggregates are used to compute a list of outlier properties and their values, as described in Section III-A. The value computed for the *sensitive files* factor is the subset of commit files that have any of the sensitive types/names specified in the configuration. The Evaluator uses the links between contributors and repository files to compute the *file history* factors: the commit files are parsed again to create two subset lists of files that have the commit author in their set of owners or majority contributors. To compute the *pull request* factor, each commit's list of associated PRs is parsed to identify any rejects as described in Section III-E.

The *contributor trust* factor is the most complicated to compute. As described in Section III-D, it is is based upon six "sub-factors" from each commit author's Contribution History. These basic properties (*e.g.* number of commits) are easily extracted from the data without additional computation; the Evaluator uses them to compute a binary (*untrusted* or *trusted*) value for this factor. To do so, a rule-based decision model compares sub-factor values to configurable thresholds. Next, a contributor is labeled *untrusted* if the proportion of violated rules meets another threshold. This model is used only for the trust factor, but *Anomalicious* uses the same type of rule-based decision model (described in the next Section) to consider the values for *all* factors and make the final decision for a commit. Below are the Trust Rule violation criteria:

**T1:** The contributor's GitHub username cannot be identified
**T2:** The contributor's GitHub account was created within the last *[threshold]* days (at time of analysis)
**T3:** The contributor has made ≤ *[threshold]* other commits
**T4:** The commit is the contributor's first
**T5:** The contributor has made ≥ *[threshold]* % of their commits on the same day
**T6:** The contributor authored their first commit within the past *[threshold]* days (at time of analysis)
**T7:** ≥ *[threshold]*% of the contributor's PRs were rejected

This "nested decision model" allows trust to be defined using several sub-factors, ensuring that commits from new/untrusted contributors are not automatically suspicious, and that trusted contributors are not automatically above suspicion.

### C. Decision Model

Once all the factors have been computed, their values are passed to the Decision Model. A customizable configuration file is used to determine which rules should be considered, the decision threshold (% rules violated of total), and other thresholds and settings for individual rules. Based on the configuration used, some rules will not apply for all commits; for example file history is not considered for "first commit" situations. These rules are designed to check if factor values meet a "violation condition" based on thresholds.
Below are the Decision Rule violation criteria:

**R1:** Commit has touched ≥ *[threshold]* sensitive files
**R2:** Commit is not the author's first AND ≥ *[threshold]*% of files have not been touched by the author before

---

**Commit:** *[Hash]*

**URL:** https://github.com/*[Username]*/event-stream/commit/*[Hash]*

**Authored on** 2018-09-09 at 08:07:49 by *[Author Name]*
**Committed on** 2018-09-09 at 08:07:49 by *[Committer Name]*
**Commit Message**: add flat map

This commit **modified 4 files**.

**33.33% of Rules were Violated**

*[Author Name]* is **TRUSTED**

The commit **changed 2 potentially 'sensitive' files**:
   package-lock.json - MODIFY - commit proportion: 100.00%
   package.json - MODIFY - commit proportion: 5.26%

**Several properties of this commit aren't typical for** *[Author Name]*:
   3 Files Modified - *[Author Name]***'s average** is 1.31
   4 Files in Commit - *[Author Name]***'s average** is 1.69
   3 Unique File Types - *[Author Name]***'s average** is 1.19

Fig. 3. Example Anomalicious Commit Report from `event-stream`

**R3:** Commit is not the author's first AND they are not majority contributor or owner for ≥ *[threshold]*% of files
**R4:** Commit adds an outlier *[author or repository]* number of files AND does not touch any files the author owns or is a majority contributor to
**R5:** Commit has ≥ *[threshold]* change properties that are outliers for the *[author or repository]*
**R6:** Commit author's contribution history for the repository indicates they are untrusted (See Section IV-B)
**R7:** Commit is linked to ≥ *[threshold]* rejected PRs

Once the Model has identified all violated rules, the proportion of violated rules to total rules is computed. This is compared to the *decision threshold*: if the proportion is greater than or equal to this threshold, the commit is considered *anomalicious* and the commit's violations and other attributes are added to the final report, described in the next section.

### D. Output: Commit Reports

After *Anomalicious* completes its analysis it produces a **Commit Report** detailing all violated rules and helpful contextual information for each flagged commit. Figure 3 shows a sample report, with identifiable information for the repository owner and commit author redacted for privacy.

First the commit hash, GitHub URL, contributors, dates, the commit message, and the total number of files modified in the commit are listed for context. Next, the proportion of violated decision model rules is given followed by the trust decision (and any violated trust rules) for the commit's author. For each violated rule, details are provided to help the maintainer understand *how* it was violated. In the example report, the author modified two sensitive files; the change type is given (MODIFY) alongside the author's proportion of the file's commits. Similar details are given for the outlier change properties; for this rule, the author's average value for each property is shown. We wanted the reports to explain exactly how the model decided the commit was anomalicious, and give enough details for a repository maintainer to quickly decide if action is needed. The time required to generate reports depends on the number of unique artifacts the repository has, due to the GitHub API's hourly request limit. In our experiments, the time needed to run Anomalicious on a repository ranged from approximately 30 seconds to 2 hours.

## V. EXPERIMENTS

In Section VI-B we discussed how *Anomalicious* was designed to meet our five criteria (Section I) for effective security-oriented anomalous commit detection: language agnosticism, customizability, low false-positive rate, explainability, and (most importantly) ability to detect actual malicious commits. The first three criteria are satisfied in the tool's design and implementation: *no language-dependent data* is collected, the factors and rules for the decision model are easily *customizable*, and the rationale of the decision model is clearly *explained* by the rules used. In this section, we discuss two experiments we conducted to verify that the *positive rate* and *malicious commit detection* criteria are met when *Anomalicious* is executed on real repositories.

### A. Positive Commit Rates and Patterns

Even with the best of intentions, an anomaly detector that is overly sensitive is not an effective solution to detect and prevent suspicious activity. Our first experiment evaluated *Anomalicious* when run on repositories with no *known* malicious commits. We wanted to investigate (1) the proportion of commits flagged, *i.e.* the positive rate, and (2) patterns of attributes and violated rules amongst these positives. We envision using our tool to notify repository maintainers of anomalicious commits using a comprehensive report that clearly explains why the commit was flagged, allowing them to quickly decide if action is needed. Prior studies have shown that contributors on GitHub already feel overwhelmed by the amount of notifications they receive [13], [14], [27], so it is extremely important that *Anomalicious* has a low positive rate.

*1) Data Collection:* For this experiment, we used a set of 100 repositories for NPM packages that have at least 100 commits [28]. We have set this commit restriction to better evaluate the ratio of flagged commits (positives) to the total commits per repository. We chose to use NPM packages for this experiment due to the history of *supply-chain attacks* [3] targeting packages in this registry. As we discuss in Section II, these attacks are especially problematic due to modern development's heavy dependence on third-party code. The number of commits per repository ranged from 100 to 18,603 with mean 1,567.83 and median 418.

*2) Experiment Design:* We used *Anomalicious* to analyze the commit histories of all 100 repositories. Using the anomalicious commit reports generated for each repository, we measured the proportion of flagged commits to total commits. The same configuration was used for the whole data set to fairly compare results. It was "tuned" by running the tool on the data set 8 times using different value combinations for the factor and decision model thresholds, keeping the configuration producing the lowest proportions of commits flagged per repository, shown in Table I.

TABLE I
CONFIGURATION FOR THE NPM DATASET (SECTION V-A2)

| Setting | Value |
|---|---|
| Decision Model Rule Threshold | 0.5 |
| Outlier Change Properties Threshold | 0.5 |
| Contrib. Trust: Min. Time as Contributor | 7 days |
| Contrib. Trust Rule Threshold | 0.5 |
| Contrib. Trust "Few Commits" Threshold | 0.05 |
| Contrib. Trust Same-Day Commit Threshold | 0.5 |
| Contrib. Trust Un-merged PR Threshold | 0.5 |
| File History Excluded Files | README, .gitignore |
| File History Consider First Commit to File | True |
| Exclude History for New Contribs. | True |
| Sensitive Files Threshold | 1 |
| Ownership Excluded Files | gitignore, class, md |
| Ownership Consider Major Contributors. | True |
| Ownership Majority Contrib. Threshold | 0.00 |
| Ownership Min. un-Owned/Majority Files | 0.75 |

TABLE II
RESULTS: COMMITS FLAGGED IN THE NPM DATASET (SECTION V-A3)

| % of Commits Flagged | % of the 100 Repositories |
|---|---|
| Less than 5% | 99% |
| Less than 2% | 86% |
| Less than 1% | 56% |

*3) Results:* The results of this experiment show that **Anomalicious has a low positive rate** for repositories with at least 100 commits. Table II summarizes the positive rates in terms of the proportion of total commits flagged and the data set proportion with that percentage. **For 99% of the repositories, $< 5\%$ of their total commits were flagged, and 56% of the repositories had $< 1\%$.**

*4) Case Study - Anomalicious Commits in Angular.js:* We examined reports for flagged commits to reason about the proportion of false positives and identify trends, patterns, or opportunities for improvement of our factors and rules. To this end, we conducted a qualitative examination on the set of positives for `Angular.js`, an extremely popular framework for JavaScript development. In our experiment, **Anomalicious flagged 0.38% (34) of the repository's 8,978 commits**. We reviewed each commit report to identify the violated rules and combined this with data from GitHub for context.

During analysis, we compared the commit messages to the actual changes made, using the visual "diff" for each commit available on GitHub. We made these comparisons to understand what types of changes were made and to look for potentially-suspicious inconsistencies. We observed that **the majority (89.25%) of positives for `Angular.js` made changes to the software supply chain**, *i.e.* development tools and infrastructure, rather than source code for the project itself. According to its developer guide[1], commit messages are required to follow a specific format that includes a type label. We observed use of 6 labels in 31 flagged commits: 'chore' (20), 'doc' (5), 'fix' (2), 'feat' (2), 'style' (1), and 'perf'(1).

---
[1]https://github.com/angular/angular.js/blob/master/DEVELOPERS.md

The guide says the 'chore' label should be used for "*changes to the build process or auxiliary tools and packages such as documentation generation*"[1]. Ten 'chore' commits updated or added individual dependencies, and another changed what tool the repository used for dependency *management*. The continuous integration and linting tools being used were also changed by 2 other commits. Three commits changed locale handling, 1 updated the project's license, and another *reverted* an unflagged 'chore' commit that modified documentation versioning updates, but broke deployment. The final 2 'chore' commits deleted "obsolete" config and settings files. Two unlabeled commits deleted a namespace and "unused" files.

The 'doc' labeled commits made updates to the framework's documentation and its generator, and the 'style' commit modified how documentation is formatted. The 2 'feat' commits added new functionality to the framework, and the 'perf' commit added dependencies and a new module for benchmark testing. The 2 'fix' commits removed a deprecated dependency and addressed a code bug. The initial commit to the repository was also flagged (due to the large number of files added), and had no label. Only 5 (14.7%) commits made changes to the source code of the framework: the 2 'feat' commits, the bug 'fix' commit, and the 'style' commit.

We also observed patterns in the rule violations: **85.29% of commits *modified files for the first time*, followed by 76.47% of commits *modifying sensitive files***. This was also the most common (70.59%) *pair* of violations. Twenty (58.82%) commits had a large number of *outlier change properties*. Rules related to file history were slightly less common: a majority of files in 47.06% (16) of the commits were not owned or heavily contributed to by the author, and an unusually large number of files were added in 35.29% (12) of commits when no files the author was a majority contributor or owner of were modified. Only 2 of the 22 unique commit authors were *untrusted* due to few total commits that were all authored on the same day. Each authored 1 of the 'chore' commits; both *modified sensitive files for the first time*. Every project and contributor is unique, so we cannot objectively determine if each commit was "worth" being notified of. However, as discussed in Section VI-B, the behavior and sensitivity of *Anomalicious*' rules and factors is guided by thresholds and other settings in a configuration file. A maintainer could easily adjust these to suit their needs. For example, if they were more interested in changes to the build system than changes to dependencies, the set of file types considered *sensitive* could be changed to only .xml files, or for even more granularity, individual files could be specified.

The majority of Angular.js's flagged commits made changes to development infrastructure, and the strongest indicators for these commits were *first changes to files*, *outlier change properties*, and *modifying sensitive files*. In other words, contributors were making anomalous changes to the infrastructure when they had never/seldom done so before. While we feel that these commits would be of interest to a maintainer, our analysis revealed that the commit reports lacked insight into the *intent* of each commit that would help them determine if action should be taken. *Anomalicious* considers some contextual factors (*rejected pull requests*, *file ownership*, and *contributor trust*), but these were weak indicators *for this repository*. Our takeaway was that ***Anomalicious*' contextual factors could be adjusted and extended to more effectively reason about *intent***. The thresholds of the existing contextual factors should be tuned for individual repositories, and the decision model could use a higher threshold for rule violations or require the presence of at least 1 contextual factor. As we reviewed commit messages and other information on each commit's GitHub page, we noticed 50% of them referenced a merged pull request or an issue - this evidence of "peer review" was very helpful to understand the intent of suspicious commits. In other words, **considering positive *and* negative factors could improve evaluation of intent.**

### B. Detecting Known Malicious Commits

In this section we discuss our second experiment evaluating *Anomalicious*' ability to detect malicious commits, our final criterion for *security-oriented* anomaly detection.

*1) **Data Collection**:* To conduct this experiment, we needed a data set of repositories with at least one *known* malicious commit, for which the relevant data is still available on GitHub. There are very few such repositories, but Table III describes the 15 we found that fit this criterion. The results of our experiment are also listed in this table, and discussed in Section V-B3. The number of commits per repository ranges from 5 to 1,830 with a mean 175.6 and median 13. Due to our use of factors that look for statistical outliers in distributions, we had to set a requirement for the minimum number of commits a repository must have. After testing this factor on repositories with 1 to 100 commits, we found at least 5 commits would suffice. Thirteen of these repositories were infected with the *Octopus Scanner Malware*. As noted in the disclosure report [19], this malware affects NetBeans projects; it alters build files and injects a file into the nbproject directory called cache.dat. To find repositories infected by this malware, we conducted a simple GitHub search for this file and manually analyzed each result to confirm other characteristics of the malware were present, *i.e.* the same changes made to the nbproject/build-impl.xml.

We also identified 2 repositories affected by other malware whose disclosures were familiar to the authors. The event-stream repository fell victim to an infamous backdoor injection by a trusted contributor who was given elevated privileges by the repository owner. The malicious actor added a dependency to the project's package.json file for a malicious package called flatstream-map. minimap is an open source code-preview package for the Atom editor. A few years ago, an employee of the code-completion company Kite authored and pushed a commit injecting advertising popups. When the community noticed this [29], a heated discussion erupted as users were outraged at Kite's takeover tactics, which had also been used in other packages [30].

TABLE III
DATASET SUMMARY (SECTION V-B1) AND RESULTS (SECTION V-B3) FOR MALICIOUS COMMIT DETECTION EXPERIMENT

| Repository | Contribs | Files | Commits | Maliciousness | Found? | Flagged | Proportion |
|---|---|---|---|---|---|---|---|
| minimap | 41 | 112 | 1662 | Backdoor Injection | No | 72 | 4.33% |
| event-stream | 37 | 29 | 291 | Backdoor Injection | **Yes** | 11 | 3.78% |
| pacman-java_ia | 10 | 266 | 119 | Octopus Scanner | **Yes** | 10 | 8.40% |
| SuperMario-FR- | 5 | 180 | 98 | Octopus Scanner | **Yes** | 3 | 3.06% |
| VehicleRentalSystem | 4 | 72 | 77 | Octopus Scanner | **Yes** | 5 | 6.49% |
| KeseQul-Desktop-Alpha | 2 | 205 | 30 | Octopus Scanner | No | 4 | 13.33% |
| BdProyecto | 3 | 19 | 15 | Octopus Scanner | **Yes** | 2 | 13.33% |
| Punto-de-Venta | 3 | 444 | 12 | Octopus Scanner | No | 1 | 8.33% |
| Snake | 1 | 21 | 9 | Octopus Scanner | **Yes** | 1 | 11.11% |
| ProyectoFiguras | 2 | 21 | 8 | Octopus Scanner | **Yes** | 1 | 12.50% |
| Secuencia-Numerica | 4 | 23 | 7 | Octopus Scanner | **Yes** | 4 | 57.14% |
| JavaPacman | 2 | 238 | 5 | Octopus Scanner | No | 0 | 0.00% |
| V2Mp3Player | 1 | 33 | 5 | Octopus Scanner | No | 0 | 0.00% |
| RatingVoteEPITECH | 2 | 38 | 5 | Octopus Scanner | No | 0 | 0.00% |
| College-GPA-Calculator | 1 | 25 | 4 | Octopus Scanner | No | 0 | 0.00% |

*2) Experiment Design:* The core design of this experiment is the same as the first experiment (Section V-A2). To "tune" the configuration settings for this data set, we used a procedure similar to that used in the first experiment: several experiments were run with different threshold values. For this data set, we chose a configuration producing results that balanced the number of repositories whose malicious commit was found with the total number of commits flagged per repository. Only three values were different from the configuration used for the first experiment shown in Table I: the decision model threshold (0.33), the outlier change property threshold (0.4), and the majority contributor threshold (0.25).

*3) Results:* Table III shows the results of this experiment; with such a small data set, we consider them to be *promising* but *preliminary*. In total, **Anomalicious identified the malicious commit in 53.33% (8) of the 15 repositories.** Each row in the table reports results for an individual repository; the yes/no values in the "Found?" column indicate whether the malicious commit was detected, and the "Flagged" column reports the number of commits that were labeled as anomalicious. Similar to our first experiment (Section V-A3), we evaluated positive rates. We found that the proportion of flagged-to-total commits is higher for this data set, but we note that there is a much smaller range in its commit distribution: only 3 (20%) of these repositories have over 100 commits. Table III shows that for two out of the eight "hit" repositories, *only* the malicious commit was flagged, the others had at least one other flagged commit. Four of the seven repositories whose malicious commit was not detected had zero flagged commits.

*4) Case Study:* Next, we examined 39 commit reports from 3 groups: 7 *undetected* malicious commits, 8 *detected* malicious commits, and 24 other positives from 3 repositories. Our analysis procedure was the same used in Section V-A4.

Each of the 8 *detected* malicious commits violated the *sensitive files* rule, and the second most common violation (5 commits) was *outlier change properties*. None of the commits violated the rule about not being a major contributor or owner to 75% or more of the commit's files, but 3 commits modified a smaller proportion of such files while also adding an unusually large amount of new files. Two of the commits had *untrusted authors*; for one it was their first commit and the other had only made 2 commits on the same day. While 7 of these repositories were victims of the Octopus Scanner malware [19], the backdoor injection in `event-stream` was also detected by the same set of violations (*sensitive files*, *outlier properties*) - supporting our claim *Anomalicious* is not restricted to specific malware.

Six of the seven *undetected* malicious commits also violated the *sensitive files* rule; two commits also had *outlier change properties*, but not enough to meet the threshold (40%) for that rule. Six commits added most of their files, but the number of new files was not an outlier for the authors. In the seventh case, the author owned the majority of files modified. All authors were trusted. These characteristics indicate that additional factors are needed to more accurately identify malicious commits from top contributors.

We chose 3 repositories for our analysis of "other" positives; the malicious commit was been detected in all three, but they differ in their number of commits and positive rates. The first two, `pacman-java_ia` and `SuperMario-FR-` were infected by the Octopus Scanner Malware and the third (`event-stream`) was victim to a backdoor injection. Ten (8.40%) of `pacman-java_ia`'s 117 commits were flagged; 9 were not previously known to be malicious. Two actually *removed* the malicious files and were flagged because the files were *sensitive*, it was the authors *first time modifying* them, and they were *untrusted* due to few commits that were all made on the same day. Four commits were refactorings that *modified sensitive files* and made *first file changes*; three modified build and settings files and the fourth renamed a package and had *outlier change properties*. **We observed suspicious behavior in three of `pacman-java_ia`'s commits that could not be explained by any contextual information**. One commit message claims that a `.jar` file was deleted and others were rebuilt, but actually two `.jars` were renamed and none were deleted. The renaming itself was also suspicious: "Pacman" was changed to "Piacman" in both cases. *Anomalicious* flagged this commit for *modifying sensitive files* (the `.jars`) *for the first time*. The author of another commit added new *sensitive configuration files*, and in a *first-time change* to a code file, listed themselves as a file author while only adding 1 comment. There was already an author listed, but they were not a contributor to `pacman-java_ia` implying the file was "borrowed". The third commit was flagged because *sensitive files* were *added when no owned or majority contributor files were modified*. The author of this commit actually adds credit to the unknown person in the README, and added a `.md` file containing Java code that mentions them again. They also added a new `.jar` file not mentioned in the commit message.

Three out of 98 commits (3.06%) were flagged for `SuperMario-FR-`. The first was known to be malicious, and followed the common pattern of *modifying sensitive files*, *first file changes*, and *outlier change properties*. The second removed the malware and the third was the repository's initial commit. Because *Anomalicious* considers commit history as a whole, this commit was flagged for *outlier change properties* and *adding an unusual number of files without modifying any owned/majority contributor files*. To change this behavior, the tool could be modified to only consider commits older than the commit-under-analysis. Finally, we analyzed the 11 commits (3.78% of 291) flagged for `event-stream`. The rationale for flagging these commits aligned with our earlier analysis of NPM repositories in Section V-A4. Namely, all of the commits performed some refactoring or added package dependencies, but the contributors lack of prior history with these files triggered the flags. Only 1 of these commits did not *modify a sensitive file*, 6 *modified a large proportion of files they did not own or majorly contribute to*, and 5 had *outlier change properties*. Aside from the known-malicious commit, we observed no other suspicious behavior in this repository.

## VI. DISCUSSION

### A. Contribution and Novelty

The main contribution of this paper is to show that *it is possible to detect malicious commits with high accuracy using security-focused anomaly detection of only commit-related metadata* that can easily be retrieved from GitHub. In contrast, prior work on anomalous commit detection [12], [14]–[17], [31] did not consider security. Moreover, prior work on malware detection [32], [33] is based on the detection of low-level detailed malware signatures or behaviors (*e.g.* API calls), not on behavioral factors from high-level metadata. Using precise malware signatures, these tools can detect malware with low false positives, but they are also restricted to known malware signatures – they cannot detect any other form of malicious commits. *Anomalicious* is different as it can detect malicious commits using only metadata, and attempts to abstract and generalize malicious commit patterns. Thus, our approach can be used to detect previously-unknown malware signatures, but at the cost of possibly more false positives.

Designing a good rule-based model for malicious commit detection is hard due to several reasons. The main reason is simply the lack of public data available; repositories infected with malware are usually either deleted or have the malicious commits expunged from their history in order not to scare potential users. Moreover, a simple model based on contributor-trust only is insufficient due to the prevalence of account hijacking for introducing malware in OSS repositories; it is then necessary to detect anomalous commits from trusted contributors in order to handle account-hijacking cases. Sometimes, the malicious commit is the first commit in the commit history: this was observed for seven repositories in the Malicious Commit data set. An existing repository was copied/cloned, infected with malware, and then used to start a new repository; in such cases, analysis of the previous commit history is impossible, and code ownership metrics are useless and even misleading/dangerous. Designing a single model that can handle all such scenarios is challenging.

### B. Satisfying Our Criteria for Effective Anomaly Detection

In Section I we outlined a set of criteria that any effective anomalous commit detector should meet: language-agnostic, customizable, explainable, and low false positives. We also specified an extra criterion for *security-oriented* anomalous commit detectors - able to identify malicious commits. These criteria served as foundational requirements in the design and construction of the *Anomalicious* tool.

To ensure **language agnosticism**, none of the factors (Section III-D) considered by the decision model depend on the contents of source code artifacts. Instead, only data that can be easily extracted from a local commit log or retrieved publicly from the GitHub API is used. *Anomalicious* is also **customizable**. Users can easily modify the Data Pipeline (Section IV-A) to collect any additional data available from the GitHub API or the local repository. They can also modify the settings file used by the Evaluator and Decision Model to adjust the number of commits being flagged, and achieve a **low positive rate**, as shown in Section V-A. When a commit is flagged, it is important for the reason to be **explainable** so that a repository maintainer can easily understand *why* the commit was flagged and quickly decide whether it is worth taking further action. To support this criteria, *Anomalicious* outputs a report (*e.g.* Figure 3) on each run that provides for each flagged commit (1) contextual details (*e.g.* commit message), (2) an enumeration of violated rules, and (3) quantitative description of how each rule was violated. Finally, *Anomalicious* is a security-oriented anomaly detector: it is able to **detect malicious commits** as shown in Section V-B.

### C. Threats to Validity

In Section V-B, we showed that our approach was able to successfully detect 8 (53%) of 15 known malicious commits with relatively few total positives among more than 2,000 commits. Our rule-based model can be further refined, as discussed earlier, which we expect will further increase the detection rate and reduce the positive rate. Overall the results are encouraging but limited due to the lack of available data for malicious commits. Many times when a malicious commit is detected the commit is purged from history. The main obstacle to further improvements in this line of research is the lack of training data available with known malicious commits in OSS repositories. We encourage readers who are aware of any examples to tweet them to us at @saintesmsr #anomalicious.

## VII. CONCLUSION

Supply-chain attacks target software development ecosystems and package repositories and have increased over the past years. We presented the *Anomalicious* tool that identified 53.33% of malicious commits in 15 malware-infested repositories, while flagging less than 5% of commits for 99% of repositories in a large-scale experiment with NPM packages. Our future work will focus on adding contextual factors to the tool. We also plan to add a mechanism to automatically adjust thresholds based on maintainer's feedback. Another direction for future work is to work on approaches to detect and prevent account hijacking and typo-squatting attacks.

**Acknowledgements.** We thank Oege de Moor, Bas Alberts, Adam Baldwin, and Ravi Gadhia from GitHub for encouraging us to pursue this line of research, for their help in collecting OSS malware datasets, and for their thoughtful feedback.